\documentstyle[a4,12pt]{article}

\begin{document}

\begin{titlepage}
\title{\normalsize\begin{flushright}nlin-sys/9702006 \\
                                    THU-97/05\\
                                    Februari 1997
                  \end{flushright}\vspace{5ex}
\Large\bf The Elliptic Billiard: Subtleties of Separability}

\author{R van Zon\thanks{Corresponding author. Email: R.vanZon@fys.ruu.nl}
\and Th W Ruijgrok \and
{\it Institute for Theoretical Physics, University of Utrecht} \and
{\it Princetonplein 5, Postbus 80.006 3508 TA Utrecht.} }

\date{Januari 31, 1997}

\maketitle\thispagestyle{empty}

\begin{abstract}

Some of the subtleties of the integrability of the elliptic quantum
billiard are discussed. A well known classical constant of the motion has
in the quantum case an ill-defined commutator with the Hamiltonian. It is
shown how this problem can be solved. A geometric picture is given
revealing why levels of a separable system cross. It is shown that the
repulsions found by Ayant and Arvieu are computational effects and that
the method used by Traiber {\em et al} is related to the present picture
which explains the crossings they find. An asymptotic formula for the
energy-levels is derived and it is found that the statistical quantities
of the spectrum $P(s)$ and $\bar{\Delta}(L)$ have the form expected for
an integrable system.\\

Quelques-unes des subtilit\'es de l'int\'egrabilit\'e du billard
elliptique quantique sont discut\'ees.  Dans le cas quantique, le
commutateur avec l'Hamiltonien d'une certaine quantit\'e connue pour
\~etre une constante du mouvement en m\'echanique classique, s'av\`ere mal
d\'efini. On d\'emontre ici comment on peut r\'esoudre ce probl\`eme.
Pour expliquer que les niveaux d'\'energie d'un syt\`eme s\'eparable se
croisent, une repr\'esentation g\'eom\'etrique est utilis\'ee. On montre
que les r\'epulsions trouv\'ees par Ayant et Arvieu sont des effets
num\'eriques et que la m\'ethode employ\'ee par Traiber {\em et al} est
li\'ee \`a la repr\'esentation g\'eom\'etrique consid\'er\'ee ici, ce qui
explique qu'ils trouvent aussi un croisement des niveaux d'\'energie.  Une
expression asymptotique est deriv\'ee pour les niveaux d'\'energie et les
quantit\'es statistiques du spectre $P(s)$ et $\bar{\Delta}(L)$ sont
obtenues. Ils ont la forme pr\'edite pour des syst\`emes int\'egrables.

\end{abstract}

\end{titlepage}

\pagenumbering{arabic}
\setcounter{page}{1}

\newcommand{\nameBerryTabor}{Berryand Tabor}
\newcommand{\nameAyant}{Ayant and Arvieu}
\newcommand{\nameTraiber}{Traiber{\em et al}}
\newcommand{\nameMcLachlan}{McLachlan}
\newcommand{\nameArscott}{Arscott}
\newcommand{\dd}[2]{\partial_{#2}{#1}}
\newcommand{\ddsqr}[2]{\partial^2_{#2}{#1}}

\section{Introduction}

Although non-relativistic quantum mechanics is a well understood theory,
about two decades ago a question arose which is still not completely
answered. We know that chaos in classical mechanics is due to nonlinear
terms in the equations of motion. The Schr\"odinger equation is linear, so
there should be no quantum chaos. But classical mechanics is supposed to
be some limit of quantum mechanics, so what is the equivalent of chaos in
quantum mechanics? By now quite some theory has been developed to answer
that question\cite{LesHouches}. The presence of chaos can be seen in the
spectrum of the Hamiltonian and its statistical properties. On varying a
parameter~$\epsilon$ of the system, two levels could approach one another.
In an integrable system, they will continue to approach and cross when
$\epsilon$ is changed further, but in nonintegrable systems, the levels
will avoid crossing: they repel.  Much research is being done on this
topic of 'Quantum Chaos'\cite{Casati}.  The assumptions underlying these
(and other) predictions are not linked rigorously to the integrable and
nonintegrable nature although in most cases they seem to hold.  Usually,
one investigates chaotic systems and determines the statistical properties
of the spectrum. Seldom an integrable system is considered, even though
such systems are not as trivial as one might expect.

In this article we look at the elliptic quantum billiard. This billiard is
often taken as a reference system to some nonintegrable
variants\cite{Sieber,Berry}, and its integrability is taken for granted.
An extensive semiclassical survey, as well as numerical solutions to the
exact eigenvalue problem, can be found in \cite{Waalkens}. We take a
closer look at the subtleties of the integrability of this billiard. The
existence of the second conserved quantity will be investigated in a
limiting scheme, involving a larger class of separable systems.  Level
crossing will be investigated and two statistical properties of the
spectrum, namely the distribution of level spacings $P(s)$ and the
rigidity $\bar{\Delta}(L)$ \cite{Mehta,BerryAgain} are used to establish
whether the system is integrable.

\section{The Elliptic Billiard}

\label{sub:billiard}

The elliptic billiard is defined as a particle moving in a two dimensional
potential well with an elliptic boundary. Classically, this system has a
second constant of motion: the product of the angular momentum $l_1$ with
respect to one focal point and the angular momentum $l_2$ with respect to
the other focal point\cite{Sieber,Berry,Waalkens,Zhang}.  This quantity
has the same value before and after a collision of the particle with the
wall, as well, of course, as during its rectilinear motion. This means
that the system is integrable, but there are some subtle points that have
not been noticed in the literature.

We formulate the problem in elliptic coordinates:
\begin{eqnarray*}
 x&=&f \cosh z \cos \theta \\
 y&=&f \sinh z \sin \theta,
\end{eqnarray*}
so that the focal points are at $(-f,0)$ and $(f,0)$. The limit to
circular coordinates can be obtained by putting $r=\frac{1}{2}f\exp(z)$,
and letting $f$ tend to zero while $r$ remains finite. Defining
\[
	M(z,\theta) = \cosh^2 z - \cos^2\theta ,
\]
$\cal H$ and ${\cal L}\equiv (l_1\;l_2 + l_2\; l_1)/2$ take the form
\[
	{\cal H} =\frac{1}{2mf^2M(z,\theta)}(p_{z}^2+p_{\theta}^2)+ V(z,\theta)
\] and \[
	{\cal L} =\frac{1}{M(z,\theta)}(\sinh^2 z\:p_{\theta}^2 -
\sin^2\theta\:p_{z}^2) .
\]
where $p_z=-i\hbar\dd{}{z}$ and $p_{\theta}=-i\hbar\dd{}{\theta}$. The
potential $V(z,\theta)$ for the billiard is zero for $z<z_{b}$ and
infinite for $z>z_{b}$. The eccentricity of the elliptic boundary is
$\epsilon=1/\cosh z_b$.  For a conserved quantity, the commutator with
${\cal H}$ should be zero. The commutator of $\cal H$ and ${\cal L}$ is
\begin{eqnarray*}
  [{\cal H},{\cal L}]   &=&
-\frac{\hbar\sin^2\theta}{M(z,\theta)} (\hbar\ddsqr{V}{z}
+ 2i\:p_{z} \dd{V}{z})
\\
 &&  +  \frac{\hbar \sinh^2 z}{M(z,\theta)} (\hbar \ddsqr{V}{\theta}
      + 2i\:p_{\theta} \dd{V}{\theta}).
\end{eqnarray*}
In our billiard, $\dd{V}{z}\rightarrow \infty$ at the boundary, making the
expression ill-defined. The problem is that ${\cal L}$ is not properly
defined on the Hilbert space of functions that are zero on the elliptic
boundary. The result is that we cannot tell whether ${\cal L}$ is
conserved or not.  In an article by \nameAyant\cite{Ayant} a few of the
lowest eigenvalues of $\cal H$ are calculated and plotted as a function of
the eccentricity, and repelling levels are seen -- a sign of
nonintegrability. This raised some confusion about the integrability, but
\nameTraiber\cite{Traiber} have shown numerically that these repulsions
are actually crossings. They admit, however, that the crossings they find
have not been established rigorously.

The eigenvalue-problem of $\cal H$ is separable in elliptic coordinates.
If we substitute
\[
	\Psi(z,\theta)=N(z)\Theta(\theta)\]\[E=\frac{2\hbar^2q}{mf^2}
\]
into the time independent Schr\"odinger equation ${\cal}H \Psi = E\Psi$ with
the
Dirichlet boundary condition $\Psi (z=z_{b}) = 0$, we get
\begin{eqnarray}
  	\ddsqr{\Theta}{\theta} + (a - 2q\cos 2\theta)\:\Theta & = & 0
\label{eq:Mat1}\\
   	\ddsqr{N}{z} - (a - 2q\cosh 2z)\:N & = & 0 \label{eq:mMat1},
\end{eqnarray}
in which $a$ is a separation constant.
Because $a$ and $q$ appear in both equations the eigenvalue problem is not
easily soluble (it also raises computational problems\cite{Waalkens,Traiber}).
These equations are called the Mathieu equation and the
modified Mathieu equation, respectively. Their solutions are Mathieu
functions\cite{McLachlan,Arscott}.
Due to symmetry, we can restrict ourselves to one quadrant,
imposing Dirichlet or Neuman boundary conditions on the $x$-axis and the
$y$-axis. This gives the standard four classes of solutions. The conditions for
$\Theta$ at $\theta=0$ and for $N$ at $z=0$ are both the same as the boundary
condition on the $x$-axis. The condition for $\Theta$ at $\theta=\pi/2$
is the boundary condition on the $y$-axis.  Furthermore $N$ should satisfy the
Dirichlet condition at $z=z_b$.
If we fix $q$, there exist countable many values of $a$ for which equation
(\ref{eq:Mat1})
has a solution.
Solutions satisfying Neuman (Dirichlet)
conditions at $\theta=0$ are called $ce_m$ ($se_{m+1}$). The index $m$ runs
from zero
to infinity. If $m$ is even, the solution satisfies the Neuman condition at
$\theta=\pi/2$. If it is odd, the Dirichlet condition is satisfied.

\section{Separability}

We have an ill-defined commutator.
To be able to define it, it is necessary to use a smooth
potential.  We start by making an Ansatz for a conserved quantity $Z$ in
the classical system of the form $Z={\cal L}+2mf^2\:Y(z,\theta)$. We
demand that
\[
	\dot{Z}=\frac{p_z(\dd{Y}{z}
+\dd{V}{z}\sin^2\theta)
+p_{\theta}(\dd{Y}{\theta}
-\dd{V}{\theta}\sinh^2 z)}{M(z,\theta)/2}
\]
be zero for all $(p_z,p_{\theta})$. From $\dd{}{z}\dd{}{\theta}Y=\dd{}{\theta}
\dd{}{z}Y$ we find that $V$ has to be of the special form
\[
	V(z,\theta) = \frac{V_{1}(z) + V_{2}(\theta)}{M(z,\theta)}
\]
This is the
class of separable systems\cite{Hietarinta,Whittaker} of which
the elliptic billiard is a limiting case. $Y$ is given by
\[
	Y(z,\theta) = \frac{V_{2}(\theta)\:\sinh^2 z -
V_{1}(z)\:\sin^2\theta}{M(z,\theta)}
\]
It can be shown that $[{\cal H},Z]=0$.
In the limit of the elliptic billiard, $V_2\equiv0$ and $V_1$
is taken to be zero inside the ellipse
and infinite outside. Then $Y$ is equal to
$V$ and will give the same boundary conditions for ${\cal L}$ as we
had for $\cal H$. In this way ${\cal L}$ will be an operator on the correct
Hilbert space, on which
we can consider ${\cal L}$ to be conserved.
The eigenvalue problem of ${\cal L}$ is equivalent to that of
${\cal H}$: we end up with the same equations (\ref{eq:Mat1}) and
(\ref{eq:mMat1}). The eigenvalues of ${\cal L}$ are
given by $(a-2q)\hbar^2$. This equivalence also means that ${\cal L}$ is of
no help to find the general solution.

There are only four types of billiards in two dimensions which have a
second constant of motion which is quadratic in the
momenta\cite{Hietarinta} and have non-complex
Hamiltonians. They correspond to rectangles, circles, ellipses and
parabolae. The parabolic billiard, which has a boundary composed of two
opposite parabolae with the same focal point, also has the subtleties of
coupled separated equations like the equations (\ref{eq:Mat1}) and
(\ref{eq:mMat1}) and an ill-defined commutator of a classically conserved
quantity with the Hamiltonian, which can also be fixed in a limiting
procedure.

\section{Characteristic Curves}

\label{sub:charcurve}

It is possible to use the separability of the system to explain why crossings
occur. For that we need to view equation (\ref{eq:Mat1}) as an eigenvalue
problem, with $a$ the eigenvalue, and $q$ some parameter. This boundary value
problem is of the Sturm-Liouville type, so the spectrum contains an infinite,
countable number of only simple eigenvalues bounded from below\cite{Sagan}. We
denote these eigenvalues by $a_m(q)$, where $m$ is the same
index as in section~\ref{sub:billiard} and $q$ indicates the dependence of the
eigenvalue on the parameter $q$. From the simplicity of the eigenvalues it
follows that they depend at least piecewise continuously on $q$.
Overall continuity can be deduced by performing a small rotation
$(a',q')=R_{\phi}(a,q)$ in equation (\ref{eq:Mat1}) with $R_{\phi}$ a rotation
over an arbitrary but small enough angle $\phi$. This gives again a
Sturm-Liouville problem, so in the rotated frame, $a_m'(q')$ has to be
piecewise continuous too, and $a_m(q)$ cannot be discontinuous.  Equation
(\ref{eq:mMat1}) can also be seen as an eigenvalue problem of the
Sturm-Liouville type, but with $q$ as the eigenvalue and $a$ as the parameter.
We denote the eigenvalues of this problem with $q_r(a)$, where the index $r$
runs from one to infinity. The $q_r(a)$ can also be seen to be continuous.

We can consider the graphs of the eigenvalues $a_m(q)$ as a set of lines in the
$(q,a)$-plane that do not intersect and we call those the $a$-curves. The same
picture can be used for the graphs of $q_r(a)$, which are the $q$-curves. Since
the values of $q$ and $a$ in the two equations have to agree, a solution to the
problem exists for every intersection point of the two sets of curves. The
values of $m$ and $r$ can be considered the quantum numbers of that solution.
We determined some of the lower ones of these so-called characteristic curves
numerically, using a discretization of equations (\ref{eq:Mat1}) and
(\ref{eq:mMat1}) and applying the
QL-algorithm on the resulting tri-diagonal matrices\cite{Recipes}. For
equation (\ref{eq:mMat1}) we took the boundary at $z_b=2$, corresponding to
an eccentricity $\epsilon$ of $1/(\cosh 2)$.
The results are plotted in figure~1.
The eigenvalue of the
Hamiltonian is proportional to the $q$-value, i.e. the projection of the
intersections of the $a$- and $q$-curves on the $q$-axis. If two points are
close together in projection on the $q$-axis, this does not mean that they are
close in the $(q,a)$-plane. When $\epsilon$ is changed, the $q$-curves shift
and the intersection points move. The projections of two points can move
towards each other, but that does not in general correspond to approaching
points or any other special case in the $(q,a)$-plane, so they will continue to
move in the same direction when $\epsilon$ is changed further. Thus they will
cross.

We can now understand the different results of \nameAyant\cite{Ayant} and
\nameTraiber\cite{Traiber}. \nameTraiber\cite{Traiber} have used an algorithm
which enables them to
calculate the $a$ value for given $q$ numerically, which are in effect the
$a$-curves.
Via a kind of Newton-Raphson
procedure they find the eigenvalues $q$ of the modified Mathieu equation. From
the above discussion, it is no surprise that in their figure the levels cross.
\nameAyant\cite{Ayant} did not obtain the eigenvalues one by one.
They choose a basis of the Hilbert space to turn the eigenvalue problem for
$\cal H$ into that of a matrix. Truncation of this matrix
gives a finite one, of which the eigenvalues can be calculated
numerically. Due to roundoff errors, a diagonalization routine can gives
spurious
repulsions. \nameAyant\cite{Ayant} do not say what kind of diagonalization
method they
used. As is shown in figure~2, using a method that can handle degeneracies
(first applying the Householder
method to get a tri-diagonal matrix, then applying the $QL$-algorithm
\cite{Recipes}), one finds the correct crossings that were also found by
\nameTraiber\cite{Traiber} in a different way. The matrix
size was 98$\times$98 and $\mu=1/\sqrt{1-\epsilon^2}$.

\section{Asymptotic Results}

According to current theory\cite{Mehta},
Random Matrix Theory can be used for nonintegrable systems.
One finds
that $P(s)=\frac{\pi}{2}\: s\: e^{-\frac{\pi}{4}s^2}$
and that $\bar{\Delta}(L)$ grows logarithmically
with $L$ in the `Gaussian Orthogonal Ensemble'. The fact
that $P(0)=0$ is a sign of level repulsion.
For integrable systems one
expects that $P(s) = e^{-s}$, which
is the distribution of level spacings in the case where the levels are
Poissonian distributed, and that $\bar{\Delta}(L)$ grows as $L/15$, for
nondegenerate levels, up to a
saturation point beyond which $\bar{\Delta}(L)$ remains
constant\cite{BerryAgain}.
A reliable calculation of $P(s)$ and $\bar{\Delta}(L)$ requires many
energy levels. We will use an asymptotic approach to
calculate the high energy eigenvalues. We
follow the Horn-Jeffreys method as in \nameMcLachlan\cite{McLachlan} and
\nameArscott\cite{Arscott}. We write $a(q)$ as an asymptotic expansion in
powers of $k=\sqrt{q}$: \[
a  =  -2k^2 + 2(2m+1) k + \alpha_0 + \sum_{i=1}^{\infty} \alpha_{i} k^{-i}
\]
The asymptotic form of the Mathieu equation can be written as the equation
for the harmonic oscillator, hence the integer constant $m$. This $m$ is
the same index as before. This $a$
is asymptotically on the $a_m$-curves corresponding to the solutions
$ce_m$ and $se_{m+1}$. For the expansion of $\Theta$ we use
\[
\Theta(\theta) \sim e^{k\chi(\theta)} \zeta(\theta)
[1+\sum_{i=1}^{\infty} k^{-i}f_{i}(\theta) ]
\]
These expressions are substituted into equation (\ref{eq:Mat1}) and terms of
equal
power of $k$ are equated. There are two independent solutions. The first
one is given by
\begin{eqnarray}
 \zeta(\theta)   &=& [\cos \theta \:\tan^{2m+1} (\theta/2 + \pi/4)]^{-1/2}
\nonumber \\
 \chi(\theta)    &=& 2 \sin \theta
\nonumber \\
 f_{i+1}(\theta) &=& - \int^\theta
\frac{\ddsqr{(f_{i}\zeta)}{\theta'}
                       + \zeta\:\sum_{j=0}^{i} \alpha_{j} f_{i-j}}
             { 4 \zeta \cos \theta'} d\theta'
\label{eq:freccur}
\end{eqnarray}
where, by definition, $f_0\equiv 1$.
In \cite{McLachlan} only the terms up to $f_0$ are
included to
find eigenvalues.
The spectrum that is found is equivalent to a two dimensional harmonic
oscillator. \nameBerryTabor\cite{BerryTabor} have calculated $P(s)$ for this
system. For some ratios of the
frequencies, $P(s)$ is not defined.  For other ratios, $P(s)$ shows some
peeked behavior, not a $e^{-s}$ behavior. They also showed that $P(s)$ can
approach $e^{-s}$ again when the system is perturbed.
Including $f_1$ could have the same effect.
{}From equation (\ref{eq:freccur}) we find
\[
      f_{1}(\theta)  = \frac{1}{8}
\left[\frac{-(m^2+m+1)\sin \theta + 2m+1}{\cos^2 \theta} \right.
\]\[
- \left.(m^2+m+1/2+2\alpha_0) \log\tan (\frac{\theta}{2}+\frac{\pi}{4}) \right]
\]
In order to obtain periodic solution we have to set the logarithmic term
equal to zero, so $\alpha_0=-(2m^2+2m+1)/4$. This is the
general strategy to obtain the $\alpha_i$'s. By induction from
(\ref{eq:freccur})
the general form of $f_i$ can
be seen to be
\[
  f_{i}(\theta) = \sum_{j=1}^{i} \frac{b^{(i)}_{j} +\: a^{(i)}_{j}\sin
\theta}{\cos^{2j} \theta}
\]
The second independent solution of equation (\ref{eq:Mat1}) is found by
substituting
$-\theta$ for $\theta$. For $ce$-type solutions, the boundary condition at
$\theta=0$ can be fulfilled
using $ce_m\propto\Theta(\theta)+\Theta(-\theta)$. The modified Mathieu
equation
(\ref{eq:mMat1}) can be found from the standard Mathieu equation
(\ref{eq:Mat1}) by substitution of $iz$ for $\theta$. The resulting
solution is called $Ce_m(z)$. Thus
$Ce_m(z)\propto\Theta(iz)+\Theta(-iz)$ is a solutions satisfying the
condition at $z=0$. The eigenvalues
are now given by the Dirichlet boundary condition at $z=z_b$, so that
the phase $\Phi(z_b)$ of $\Theta(iz_b)$ should be
$(r+\gamma) \pi$, where $r$ is the same index as in section~\ref{sub:charcurve}
and $\gamma=\frac{1}{2}$. For $se$-type solutions, we start with
$se_m \propto \Theta(\theta)-\Theta(-\theta)$, and we find the same
requirement, but with $\gamma=0$.
The phase can be expressed in terms of $\epsilon$ and the $a_j^{(i)}$ and
$b_j^{(i)}$:
\[
      \Phi(z_{b})  \sim 2k\frac{\sqrt{1-\epsilon^2}}{\epsilon}-(2m+1)
   \arctan\sqrt{\frac{1-\epsilon}{1+\epsilon}}
   \]\[+ \arctan\left[\frac{\sqrt{1-\epsilon^2}}{\epsilon}
    \frac{\sum_{i}\sum_{j}
a_{j}^{(i)}\epsilon^{2j}k^{-i}}{1+\sum_{i}\sum_{j}
b_{j}^{(i)}\epsilon^{2j}k^{-i}}\right]
\]
which should be equal to $(r+\gamma)\pi$.
Using the form of $f_1$, we obtain the first order equation for $k$:
\begin{eqnarray}
   k& = & (r+\gamma)\:\omega_{1} + (m+1/2)\:\omega_{2}/2
\nonumber \\
&+&\frac{\omega_1}{\pi}
	\arctan \left[ \epsilon\sqrt{1-\epsilon^2}\frac{m^2+m+1}{8k+\epsilon^2
(2m+1)} \right] \label{eq:nextorder}
\end{eqnarray}
where
\begin{eqnarray*}
	\omega_1 &=& \frac{\pi\epsilon}{2\sqrt{1-\epsilon^2}}
\\
	\frac{\omega_2}{\omega_1} &=&
\frac{4}{\pi}\arctan\sqrt{\frac{1-\epsilon}{1+\epsilon}}
\end{eqnarray*}
The accuracy improves as $k$ gets larger and $\epsilon$ gets closer to
one. For $\epsilon=0$, corresponding to the circle, it is not a good
approximation.
Equation (\ref{eq:nextorder}) is a transcendental equation for $k$, to be
solved
for each pair of quantum numbers $m$ and $r$. The lowest order
eigenvalues, given by
the first two terms in (\ref{eq:nextorder}) form a set of lines in the
$(\epsilon,k)$-plane, one line for every pair $(r,m)$. Lines with
equal $m$ but different $r$ are shifted in the $k$ direction by a
multiple of
$\omega_1$, which is not zero except at $\epsilon=0$, so they will never
cross for $\epsilon>0$. But lines with different
$m$ do cross, at least in lowest order.
The correction term in equation (\ref{eq:nextorder}) can be seen to be at
most $\omega_1/2$.
This determines a band in the
$(k,\epsilon)$-plane to which the lines are certainly confined. If the lines
remain continuous when all orders are taken into account, then they have to
intersect in some point in the area where these bands overlap.  If
$k$ is
determined by $f(k,\epsilon)=0$, the implicit function theorem states that
$k(\epsilon)$ is continuous provided that $\dd{f}{k}(k,\epsilon)\neq
0$. One easily checks that for equation (\ref{eq:nextorder}) this is the
case, so the solution is continuous and crossing is inevitable.

We solved equation (\ref{eq:nextorder}) numerically,
for about 15000 levels of the $ce$-type, for even $m$. We
took the 10000 largest of those to compute $P(s)$ and $\bar{\Delta}(L)$. For
the unfolding of the spectrum\cite{Bohigas} we took for
the accumulated level density
\[
  N(k) = \frac{( k+\omega_2/2-\omega_1)^2 - (\omega_1^2+\omega_2^2)/12
}{2\omega_1\omega_2}
\]
which follows from the eigenvalues calculated to lowest order.
The results are shown in figure~3 for eccentricity $\epsilon=0.8$.
We see the expected behavior for integrable systems.
The graphs look roughly alike for all other values of $\epsilon$, although for
some values of the
eccentricity, the first correction term in equation (\ref{eq:nextorder})
cannot totally restore the $e^{-s}$ behavior, namely when
$\omega_2/\omega_1$
is a rational number $z=p/q$, which is at $\epsilon=\cos(z\pi/2)$.
This is most pronounced for ratios $z$ of $1/3$, $1/2$ and $2/3$.

\section{Conclusions}

It is possible to define a second constant of motion for
the elliptic billiard, but only as a limiting case and the boundary has to be
included into this quantity. Separability does not mean we can solve the
system but it does provide a geometric picture in which the energy
eigenvalues are projection of intersections of characteristic curves. As
the curves change continuously when the eccentricity is varied, the
energy levels will cross generically. The level repulsions found in
\nameAyant\cite{Ayant} were not correct, due to the diagonalization method
used.
\nameTraiber\cite{Traiber} effectively used the characteristic curves,
therefore the
crossing levels that we expect were found.  The separability also allows for
an asymptotic method to obtain the spectrum, which indeed gives results which
are
characteristic for integrable systems.
So the elliptic billiard turns
out to be an ordinary integrable system, despite the subtleties in the
formalism.

\section*{Acknowledgements}
We would like to thank J. Jos\'e for his encouragement and interest
in this problem, and N.G. van Kampen for useful discussions.

\newcommand{\refformA}[6]{#1 #6 {\em #3}~#4 #5}

\newcommand{\refformB}[6]{#1 #4 {\em #2}\ #6\ (#5: #3)}

\section*{Figure Captions}

\begin{description}
 \item[Figure 1] The two independent sets of characteristic curves.
 The solid curves are the $a$-curves corresponding to the solutions $se_{m+1}$,
 the dashed curves are the $q$-curves for eccentricity $\epsilon=1/(\cosh 2)$.
 \item[Figure 2] Crossing lower energy levels as a function of
 $\mu=1/\sqrt{1-\epsilon^2}$. The energy is given in units of
 $\frac{\hbar^2}{2mf^2}(\mu-\mu^{-1})$, as in \nameAyant\cite{Ayant} and
 \nameTraiber\cite{Traiber}.
 \item[Figure 3] $P(s)$ for eccentricity 0.80. The bars are
 the calculated points, the dashed line is the theoretical
 prediction for an integrable system. The inset shows $\bar{\Delta}(L)$
 for the same eccentricity. The solid line consists of calculated points,
 the dashed line is the theoretical prediction $\bar{\Delta}(L)=L/15$ for
 small $L$  for integrable (nondegenerate) systems. For large $L$ the
prediction
 is that $\bar{\Delta}(L)$ saturates.
\end{description}

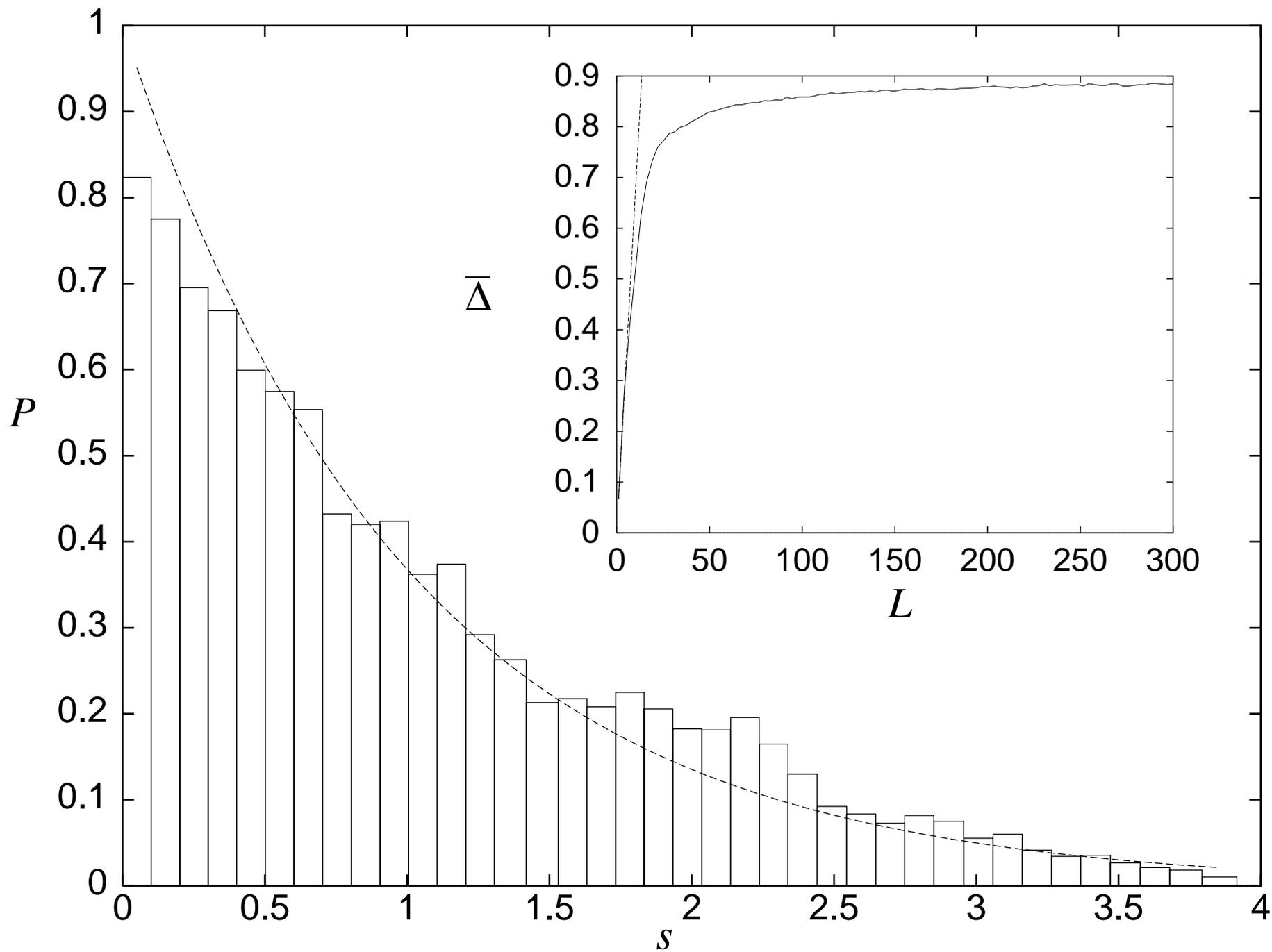
\end{document}